\documentclass[preprint,aps,prd,10pt,superscriptaddress,showpacs]{revtex4-1}
\usepackage{hyperref}
\usepackage{graphicx}
\usepackage{amsmath,amssymb}
\usepackage{subfigure,dcolumn}
\usepackage[T2A,T1]{fontenc}
\usepackage[russian,english]{babel}
\usepackage{epstopdf}
\usepackage[dvipsnames]{xcolor}
\usepackage{color}
\usepackage{multirow,booktabs}
\usepackage{lipsum}
\usepackage{lineno}
\usepackage{mathrsfs}

\usepackage{listings}
\newcommand{\figref}{Fig.~\ref}
\newcommand{\forref}[1]{Eqn.~(\ref{#1})}
\newcommand{\zbl}[1]{\textcolor{black}{#1}}

\newcommand{\zblb}[1]{\textcolor{black}{#1}}

\usepackage{xcolor}
\usepackage[normalem]{ulem} % use normalem to protect \emph
\newcommand\hll{\bgroup\markoverwith
	{\textcolor{yellow}{\rule[-.5ex]{2pt}{2.5ex}}}\ULon}

\begin{document}

\title{Constraints on cosmic-ray boosted Dark Matter in CDEX-10}

\author{Zhan-Hong Lei}
 \email{leizhh3@mail2.sysu.edu.cn}
\author{Jian Tang}%
 \email{tangjian5@mail.sysu.edu.cn}
\author{Bing-Long Zhang}
 \email{zhangblong@mail2.sysu.edu.cn, to whom all correspondence should be addressed.}
\affiliation{
School of Physics, Sun Yat-Sen University, Guangzhou 510275, China
}%
\date{\today}

\begin{abstract}
Dark matter (DM) direct detection experiments have been setting strong limits on the DM-nucleon scattering cross section at the DM mass above a few GeV, but leave large parameter space unexplored in the low mass region.
DM is likely to be scattered and boosted by relativistic cosmic rays in the expanding universe if it can generate nuclear recoils in direct detection experiments to offer observable signals.
Since low energy threshold detectors using Germanium have provided good constraints on ordinary halo GeV-scale DM, it is necessary to re-analyze 102.8 kg$\times$day data in the CDEX-10 experiment assuming that DM is boosted by cosmic rays. For the DM mass range 1 keV $<m_\chi <$ 1 MeV and the effective distance within 1 kpc, we reach an almost flat floor limit at $8.32\times10^{-30}$ cm$^2$ on spin-independent DM-nucleon scattering cross section at a 90\% confidence level. The CDEX-10 result is able to close the gap unambiguously in the parameter space between MiniBooNE and XENON1T constraints which was partially hindered by the Earth attenuation effect. \zbl{We also quantitatively calculate expected neutrino floor on searching for CRBDM in future direct detection experiments using Germanium.}
\end{abstract}

\maketitle
% \linenumbers
\section{Introduction}
%Evidence of DM in the universe, important to search for DM particles
Compelling astrophysical and cosmological evidence points to the existence of DM~\cite{Bertone:2004pz, Ade:2015xua}, yet very little is known about particle nature of DM. If DM particles could annihilate or decay into standard model particles, observations of gamma rays, anti-protons and positrons in cosmic ray telescopes and neutrinos in the related detectors will provide clues for DM properties and even its origin indirectly~\cite{Cirelli:2010xx, Gaskins:2016cha, Aguilar:2013qda, Ambrosi:2017wek}. In addition, several decades have been spent on searching for DM directly with low-background detectors in the deep underground laboratories on Earth in order to turn the DM-nucleus or DM-electron recoiled energy into light, heat and ionization signals. DM direct detection experiments intend to distinguish nuclear recoils (NRs) and electronic recoils (ERs) with the noble gas liquid detectors such as PandaX~\cite{PandaX}, Xenon1T~\cite{XENON1Tn} and LUX~\cite{LUX} using liquid Xenon, DEAP-3600~\cite{Ajaj:2019imk} and DarkSide~\cite{Agnes:2018ves} using liquid Argon, while a few of complementary experiments with crystals register DM energy deposits only, such as CoGeNT~\cite{Aalseth:2012if} and CDEX~\cite{CDEX10} using Germanium. Rapid progress has been made, but we have to be confronted with null results more often than not. As a result, stringent exclusion limits have be placed in the plane of the DM mass $m_\chi$ and the DM coupling strength $\sigma_{\chi0}$, given the weakly interacting massive particle (WIMP) as a DM candidate in the standard halo model (SHM)~\cite{TSHM,SHM}. 

In the high mass DM region, the neutrino floor caused by coherent elastic neutrino-nucleus scatterings (CE$\nu$NS) is around the corner and leads to intrinsic neutrino backgrounds for the next-generation large scale DM detectors~\cite{Strigari:2009bq, Davis:2014ama, Billard:2013qya, Strigari:2016ztv, OHare:2016pjy, Boehm:2018sux, Dutta:2019oaj}. Suffering from the energy threshold in the current detector technology, a relatively large space remains unexplored in the light DM scheme. Tremendous efforts are still underway, aiming at ultra-low threshold and large-scale detectors to reach better sensitivities. Novel experimental detection strategies and a number of new theoretical ideas to probe the low mass DM have been proposed and actively discussed very recently~\cite{Battaglieri:2017aum}, including better detection techniques to overcome the energy-recoil threshold barrier~\cite{Essig:2015cda,Hochberg_2016}, Migdal effect to observe electrons dissociated from the atom through the nuclear scattering~\cite{Ibe_2018, Bell_2020, CRESST2019, EDELWEISS, Liu:2019kzq}, the boosted DM from heavy-generation annihilations in a multi-component DM model in order to surpass the detector threshold limitation~\cite{Agashe:2014yua}. One of the interesting topics to explore low-mass DM is the cosmic ray boosted DM (CRBDM) mechanism proposed in recent studies~\cite{Yin:2018yjn,CRBDM,Ema:2018bih,JB,Dror:2019onn, Dror:2019dib, Guo:2020drq}. In the galaxy, there must be an isotropic and relativistic cosmic-ray stream~\cite{Yoon:2011aa, Strong:2007nh} consisting of protons ($\sim$86\%), helium nuclei ($\sim$11\%), some other heavier nuclei and electrons ($\sim$3\%). DM can generate nuclear recoils in direct detection experiments to offer observable signals. It is likely that DM is scattered and boosted by relativistic cosmic rays in the expanding universe. Several studies have already investigated the CRBDM and offered relatively tight constraints on light DM, including the diurnal effect~\cite{ge2020boosted}, DM-nucleon interactions propagated by various mediators~\cite{Bondarenko:2019vrb, BDMLag, Wang:2019jtk, Cho:2020mnc} and the so-called reverse direct detection method~\cite{Cappiello:2018hsu}. There are cosmological constraints from Big Bang Nucleisynthesis (BBN) and the DM relic density. Previous studies show that CRBDM with large enough cross section is expected to achieve thermal equilibrium with SM particles before the BBN epoch and would affect the prediction of BBN. Consequently, current observation of BBN excludes the DM mass below a few MeV, which still leaves a viable parameter region to be probed~\cite{Krnjaic:2019dzc}. In addition, the DM relic density puts constraints on the low mass DM, but it largely depends on model assumptions~\cite{Tsai_2013,Bhat_2020}.

As we noticed in the previous CRBDM studies, the attenuation effect results in the upper bound of the exclusion region for CRBDM. Both MiniBooNE and XENON1T provide lower bounds on CRBDM, despite the dependence of DM and cosmic ray distribution in the galaxy. The attenuation effect in XENON1T cuts in the exclusion region of CRBDM. Therefore, there exists a gap in the parameter space between MiniBooNE and XENON1T constraints~\cite{CRBDM}. Compared with other types of detector, Germanium detectors with the lower energy threshold and better energy resolution are effective means to set limits on the scattering cross section for the GeV-scale DM in the standard WIMP detection. As a consequence, the CDEX experiment located at 2.4 km underground using the point contact high purity germanium detector might be able to close the gap due to its lower level of backgrounds and lower energy threshold compared to MiniBooNE and XENON1T. Based on the CRBDM mechanism, it must be valuable to re-analyze the 102.8 kg$\times$day data of CDEX-10 experiment, and provide constraints on spin-independent DM-nucleon cross section for $m_\chi < 1$ GeV. 
%\zbl{In addition, considering that CE$\nu$NS would induce an irreducible background for future DM direct detection experiments, we also provide the neutrino floor at the low mass region for the Germanium detector.}

This article is organized as follows. In Sec.~\ref{sec:CRBDM}, we review the mechanism of CRBDM and derive the theoretical energy recoil spectrum with the Earth attenuation effect taken into account. In Sec.~\ref{sec:constraints}, we re-analyze the CDEX-10 data in the CRBDM hypothesis. Finally, we summarize and make conclusions in Sec.~\ref{sec:conclusion}.

\section{Cosmic ray boosted dark matter (CRBDM)}
\label{sec:CRBDM}
\begin{figure}[!t]
\centering
\includegraphics[width=0.8\textwidth]{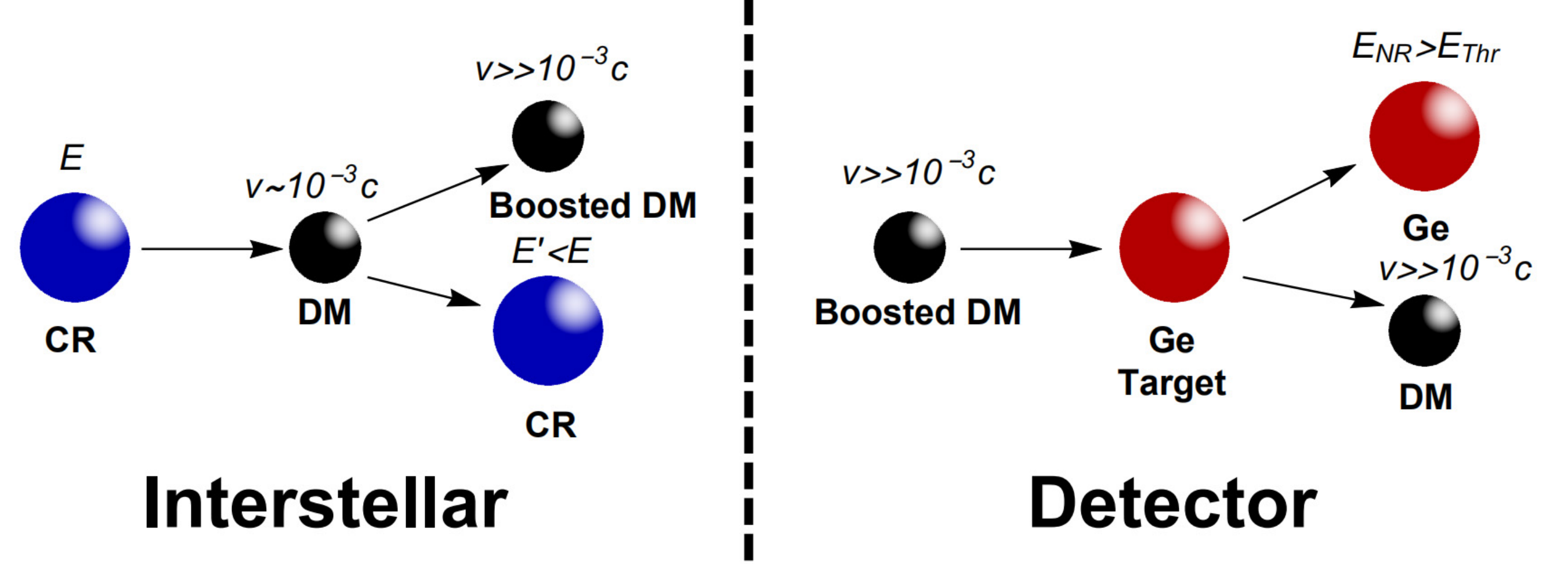}
\caption{The schematic diagram for acceleration and detection processes in a CRBDM mechanism.}
\label{fig:pic1}
\end{figure}

In a CRBDM mechanism, cosmic rays and DM exchange their energy and momentum in the process of collisions, so that the boosted DM would become almost relativistic.
Then, boosted DM particles traveling toward the detector on the Earth and scattering off the target nuclei, could generate enough recoil energy which would exceed the threshold energy. The schematic diagram to illustrate acceleration and detection processes is given in Fig.~\ref{fig:pic1}.
To study a scattering process, we need to know the information of the incident particle, the target, and the collision process.
The local density of DM near the solar system is $\rho_{\chi 0} \sim 0.4 ~\textrm{GeV/cm}^3$ according to the cosmological observations~\cite{Ade:2015xua}. 
The velocity distribution of DM in the standard halo model (SHM)~\cite{TSHM} approximates to the Maxwell-Boltzmann distribution whose the most probable velocity $v_0$ is about 220 km/s.
In the following, we will discuss the CRs and review the collision process between CRs and DM.

\subsection{Boosted DM flux from CRs}
\label{subsec:flux}

CRs might originate from some of the supernova remnant (SNR) in the Galactic Disk~\cite{Cesarsky:1980pm}.
Fermi pointed out that CRs from SNR bounce constantly between the shock surface and the magnetic field behind the wave surface.
To explain the highly isotropic distributions for energetic charged particles, the concept of CR diffusion was proposed, in which the Galactic magnetic field plays an important role in the process~\cite{Strong:2007nh}.
To calculate the differential CR flux, two main components proton and helium nucleus are considered and marked with subscript $i$ in the following formula.
With the data from PAMELA, the fitting formula of the Local Interstellar Spectra (LIS) of CRs are parameterized~\cite{spectrum2} as follows:
\begin{equation}
\label{equ::CRs_spectrum}
\frac{d\Phi_i}{dR}\times R^{2.7} =
\left\{
	\begin{aligned}
		&\sum_{j=0}^5 a_j R^j & R\leq 1~\textrm{GV}, \\
		&b+\frac{c}{R}+\frac{d_1}{d_2+R}+
		\frac{e_1}{e_2+R}+\frac{f_1}{f_2+R}+gR & R> 1~\textrm{GV},
	\end{aligned}
\right.
\end{equation}
where $R$ means the rigidity of each nucleus.
For a particle with a proton number Z, the relationship between the rigidity $R$ and the kinetic energy of CR particle $T_i$ is: $\frac{T_i}{R} = Z e$.
As we all know, the differential CR flux obeys a power law with an index $\sim -2.7$.
As shown in \forref{equ::CRs_spectrum}, $a,~b,~c,~d,~e,~f,~g$ are the numerical coefficients~\cite{spectrum,spectrum2} and we show the coefficients of proton and helium nucleus in Table~\ref{table::para}.
Considering that CRs is isotropic in the interstellar, we integrate the volume to obtain the differential CR flux $\frac{d\Phi_i}{dT_i}$: $\frac{d\Phi_i}{dT_i} = 4 \pi \frac{d\Phi_i}{dR} \frac{dR}{dT_i}$.
		
\begin{table}[!b]
\centering
\caption{\label{table::para}Parameters of the analytical fits to the proton and He LIS~\cite{spectrum}.}
\begin{tabular}{|l|c|c|c|c|c|c|c|c|}\hline
	&$a_0$ &$a_1$ &$a_2$ &$a_3$ &$a_4$ &$a_5$ &$b$ &c \\\hline
	p &94.1 &-831 &0 &16700 &-10200 &0 &10800 &8500\\\hline
	He &1.14 &0 &-118 &578 &0 &-87 &3120 &-5530\\\hline
	&$d_1$ &$d_2$ &$e_1$ &$e_2$ &$f_1$ &$f_2$ &$g$ &\\\hline
	p &-4230000 &3190 &274000 &17.4 &-39400 &0.464 &0&\\\hline
	He &3370 &1.29 &134000 &88.5 &-1170000 &861 &0.03&\\\hline
\end{tabular}
\end{table}
		
Compared with the relativistic velocity of CRs, DM at a velocity of $\sim 10^{-3}~c$ in the SHM can be safely treated at rest.
Considering the collision of DM and CRs, although the inelastic scattering occurs due to the high energy of CRs, it only contribute to a negligible portion in the DM fluxes boosted by CRs compared to elastic scattering, and the signals of $\gamma$ rays and neutrinos from inelastic scattering are beyond our topic~\cite{Guo:2020oum}. Consequently, it is reasonable for us to consider only the elastic scattering process. With the assumption that the collision is elastic and isotropic, 
we relate the kinetic energy of DM at $T_\chi$ to that of CRs at $T_i$ by energy and momentum conservation in the following:
\begin{equation}
\label{equ::relativty_relation}
	T_\chi = \frac{T_i^2 + 2m_i T_i}
	{T_i + \frac{(m_i+m_\chi)^2}{2 m_\chi}}  \frac{1 + \cos{\theta}}{2} \,,
\end{equation}
where $\theta$ is the scattering angle in the center of momentum frame, and $m_i$ is the mass for CR particle $i$.
According to \forref{equ::relativty_relation}, $T_\chi$ tends to decrease and is proportional to $m_\chi $ if $m_\chi \ll m_i$, while DM would gain more energy if $m_\chi$ is close to $m_i$.
DM gets the maximimal kinetic energy $T_\chi^\mathrm{max}(T_i)$ when the CR direction is completely reverted after such a collision ($\theta=0$).
%If $\theta$ is zero, we have back to back scatterings and DM would gain the maximum recoil energy $T_\chi^\mathrm{max}(T_i)$. 
The minimum energy of the incident particle $T_i^\mathrm{min}(T_\chi)$ is given in~\forref{equ::min}:
\begin{equation}
	\label{equ::min}
	T_i^\mathrm{min} \left(T_\chi\right) = \left(\frac{T_\chi}{2} - m_i\right)
	\left[ 1\pm \sqrt{1 + \frac{2T_\chi}{m_\chi} \frac{(m_i+m_\chi)^2}{(2m_i-T_\chi)^2}} \right]\,. 
\end{equation}
%, can be obtained by inverting the function \forref{equ::relativty_relation}.
Note that the sign $+/-$ applies to the case with $T_\chi > 2 m_i$ or $T_\chi < 2 m_i$.

When momentum transfer occurs in the scattering process, the internal structure of the nuclei would take effect.
Therefore, the cross section varies with the exchange of momentum described by the form factor $G_i$: $\sigma_{\chi i} = \sigma_{\chi i}^0 G_i^2(2m_\chi T_\chi)$,
where $\sigma_{\chi i}^0$ is the cross section for zero momentum transfer.
In the collision process with large momentum transfer, just as the case that DM is boosted by CRs, we adopt the dipole form to describe the effect of nonpoint-like structure of nucleon. Note that for the small momentum transfer, the Helm model is preferred. The dipole form factor of proton and helium is given by~\cite{Perdrisat:2006hj}:
\begin{equation}
	G_i(2m_\chi T_\chi) = \frac{1}{(1+ 2m_\chi T_\chi / \Lambda_i^2)^2}\,,
\end{equation}
where $\Lambda_i$ is inversely proportional to the charge radius, and we set $\Lambda_p = 770$ MeV for proton and $\Lambda_{He} = 410$ MeV~\cite{radius} for helium nucleus.
It is generally believed that the DM couplings to the proton and nucleon are the same, so the cross section can be written as \forref{equ::A2} with a dependence of the mass number $A$ for the incident particle $i$:
\begin{equation}
\label{equ::A2}
\sigma_{\chi i}^{0} = \sigma_{\chi 0} A^2 [\frac{m_{i} (m_{\chi}+m_{p})}{m_{p} (m_{\chi}+m_{i})}]^2\,.
\end{equation}
It is worth mentioning that the factor $A^2$ in \forref{equ::A2} is 16 for the helium nucleus and 1 for the proton, respectively. The Helium nucleus and proton make contributions at the same order of magnitude to the differential flux for DM (the helium nucleus account for 10\% of CRs).

\begin{figure}[!t]
\centering
\includegraphics[width=0.7\linewidth]{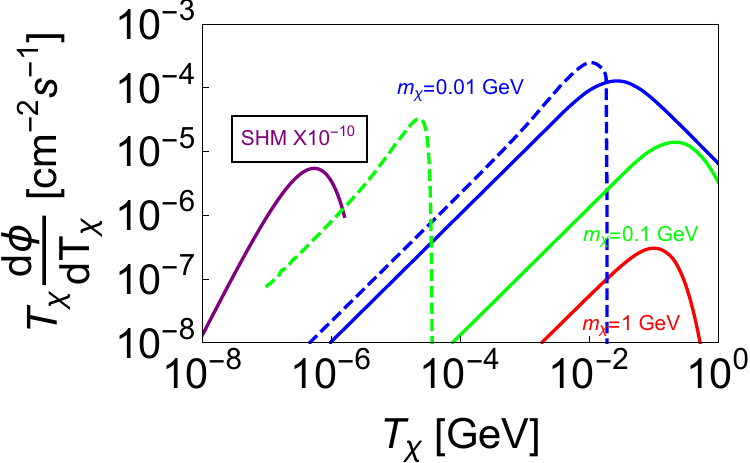}
\caption{The fluxes of CRBDM are given in solid lines for the DM mass $m_\chi = 0.01$ GeV (blue),  $m_\chi = 0.1$ GeV (green) and $m_\chi = 1$ GeV (red), assuming $\sigma_{\chi 0} = 10^{-28} ~\textrm{cm}^2$ and D$_\textrm{eff}$ =1 kpc. The purple solid line is the DM flux with $m_\chi = 1$ GeV for SHM normalized by $10^{-10}$. All dashed lines represent fluxes of CRBDM with the Earth attenuation effect taken into account.}
\label{fig:flux}
\end{figure}

Based on the above information, we start to calculate the differential flux for boosted DM.
Inside a volume $dV$, the collision rate with energy exchange $dT_i$ for CRs and $dT_\chi $ for DM is given by:
\begin{equation}
\label{equ::probability}
d\Gamma_{CR \rightarrow \chi} = \sum_i \frac{\rho_\chi}{m_\chi} \times  \frac{d \sigma_{\chi i}^0}{d T_\chi} G_i^2(2m_\chi T_\chi)  \times
			\frac{d\Phi_i}{dT_i}  dT_i dT_\chi dV \,,
\end{equation}
where $\frac{d\Phi_i}{dT_i}$ is the differential flux for CRs.
Without a loss of generality, we make an energy-independent assumption here with such a form factor untouched that energy transfer is uniformly distributed as 
$\frac{d\sigma_{\chi i}^0}{dT_\chi} = \frac{\sigma_{\chi i}^0}{T_\chi^{max}}$.
By integrating the volume and the energy $T_i$, we obtain the differential flux for the boosted DM in~\forref{equ::dmspectrum}.
In addition, the volume integral asks for the distribution of DM and CRs in the galaxy. Note that the integral needs to be divided by the factor $4 \pi d^2$ to obtain the DM flux boosted by CRs, where $d$ is the distance between the collision point and the Earth. 
We introduce the effective distance $D_{\text{eff}}$ to represent the volumn integral, into which we put the results of different assumptions on the CR spatial distribution. The CR spatial distribution is considered as the homogeneous and spherical distribution given in the reference~\cite{CRBDM} and results in $D_{\text{eff}}=1$ kpc conservatively. While other researchers assume that CRs are uniformly distributed in a cylindrical distribution with a radius $R = 10$ kpc and a half-height $h = 1$ kpc~\cite{Ema:2018bih, JB} and results in $D_{\text{eff}}=3.7$ kpc. Also, the CR distribution can be simulated with the GALPROP code~\cite{ge2020boosted}. 
Therefore, it is worth further investigations on the CR distribution in the future work and we now take two benchmark examples of $D_{\text{eff}}$ for analyses:
\begin{equation}
\label{equ::dmspectrum}
	\frac{d\Phi_\chi}{dT_\chi} =
	\sum_i \int \frac{d \Omega}{4 \pi} \int dl \int_{T_i^{min}}^{\infty}
	\frac{\rho_\chi}{m_\chi} \frac{d\sigma_{\chi i}^0}{dT_\chi} G_i^2(2m_\chi T_\chi)
	\frac{d\Phi_i}{dT_i} dT_i
	=
	D_{\text{eff}}  \frac{\rho_\mathrm{\chi 0}}{m_\chi}
	\sum_i  \sigma_{\chi i}
	\int_{T_i^\mathrm{min}}^\infty
	\frac{1}{T_\chi^\mathrm{max}(T_i)}
	\frac{d\Phi_i}{dT_i} dT_i\,,
\end{equation}
where $T_i^\mathrm{min}$ and $T_\chi^\mathrm{max}(T_i)$ are given in~\forref{equ::min} and \forref{equ::relativty_relation} with $\theta=0$.
Following the uncertainty analyses on the DM local density and the DM spatial distribution~\cite{JB}, we take 50\% uncertainty for analyses.
To further study the boosted DM, we can make a comparison between the boosted DM flux induced by CRs and the DM flux under the SHM in the same reference frame:
\begin{equation}
\label{equ::compare}
\frac{d\Phi_\chi^{SHM}}{dT_\chi} = \frac{\rho_\mathrm{\chi 0}}{m_\chi} v f(v) \frac{dv}{dT_\chi}
= \rho_\mathrm{\chi 0} m_\chi f(v) \frac{1}{(T_\chi + m_\chi)^3}\,,
\end{equation}
where $f(v)$ is the Maxwellian velocity distribution with the most probable velocity at $v_0=220 ~\textrm{km/s}$ and the galactic escape velocity at $v_{escape}=540 ~\textrm{km/s}$~\cite{DMVescape}. Here the velocity distribution in SHM is taken as follows:
\begin{equation}
f(v) = \frac{1}{k} 4 \pi v^2 e^{-v^2/v_0^2} \Theta (v-v_{escape})\,,
\end{equation}
where $k$ is the normalization factor.
The numerical results for the CRBDM fluxes are compared and shown in~\figref{fig:flux}.
As the DM density in space is fixed, the smaller DM mass results in the larger number density of DM. 
Compared with the DM flux under the SHM hypothesis, the flux intensity boosted by CRs is a small portion of the total flux.
Only a small portion of CRs lose energy, so that the CRBDM mechanism can survive in the current constraints from astrophysical experiments. It provides a clue for the reverse direct detection for DM~\cite{Cappiello:2018hsu}.

\subsection{Attenuation}
\label{subsec:attenuation}
As DM particles could collide with ordinary matter, the DM flux from space to the detector would be blocked by the atmosphere and the rock, which causes energy loss or de-acceleration for DM.
We simply estimate the mean free path for DM as $\lambda = 1/\sigma_{\chi N} n_\mathrm{N}$ where $\sigma_{\chi N}$ stands for the DM-nucleus scattering cross section.
Taking the mean mass number $\overline{A} = 20$ in the Earth crust, the mean density $\rho_{crust} = 2.7$~g/cm$^3$ in the crust~\cite{attenuation} and $\sigma_{\chi 0} = 10^{-30} $ cm$^2$ as parameters, we get the the mean free path: $\lambda  \sim  0.1 $ km.
In order to suppress backgrounds from cosmic rays, direct detection experiments are usually located deep underground. For example, China Jinping Underground Laboratory to host the CDEX experiment has the rock overburden at $z \sim 2.4$ km. Then it is time to establish a model to analyze the Earth attenuation effect and the ballistic trajectory approximation is adopted.

Under the assumption of energy independence: $\frac{d \sigma_{\chi N}}{d T_{r}} = \frac{\sigma_{\chi N}}{T_{r}^{\max }}$, we obtain the energy loss for DM crossing the crust:
\begin{equation}
\label{equ::attenuation}
	\frac{d T_\chi}{d z}=-\sum_{N} n_{N} \int_{0}^{T_{r}^{\max }} \frac{d \sigma_{\chi N}}{d T_{r}} T_{r} d T_{r}
	= -\frac{1}{2} \sum_{N} n_{N} \sigma_{\chi N} T_{r}^{\max }\,,
\end{equation}
where $T_{r}^{\max }$ is the maximal recoil energy between the DM particle and the nucleus $N$. Here the form $T_{r}^{\max}$ is similar to \forref{equ::relativty_relation} with $\theta=0$ by a replacement of $ m_\chi \rightarrow m_N$ and $m_i \rightarrow m_\chi$.
We have to solve the differential equation in~\forref{equ::attenuation} to establish the relationship between DM energy in the space and detector: $T_\chi^z = T_\chi^z (T_\chi^0)$.
However, recent studies~\cite{Xia:2021vbz,PandaX:2021kai} involve the form factor in the energy loss calculations and make necessary modifications on \forref{equ::attenuation}, which leads to a different constraint on the parameter space of DM. In the following section, we will make a comparison of these effects in the exclusion plots.
After that, we will get the attenuated DM flux spectrum as follows:
\begin{equation}
	\frac{d\Phi_{\chi}}{dT_\chi^z} = \frac{d\Phi_{\chi}}{dT_\chi^0}
		\frac{dT_\chi^0}{dT_\chi^z}\,.
\end{equation}

Adopting the nuclear composition of the Earth crust given in the reference~\cite{attenuation} and the crust thickness as 2400~m, we present the DM flux with/without the Earth attenuation effect in \figref{fig:flux}. The comparison of different line colors tells us the interesting features associated with the DM mass while the dashed/solid lines highlight the cases with/without the attenuation effect, respectively. Under the same assumption for the cosmic ray acceleration mechanism, we see the low mass DM gets boosted into higher flux intensity region. However, the high mass DM tends to be bounced towards the low kinetic energy part by rock as the Eartch attenuation effect is taken into account. 
If such a bounce on Earth for CRBDM are too strong, the DM particle might not be able to reach the detector. 
The energy loss of DM particles crossing the earth crust depends on the maximal recoil energy, which also depends on the DM mass. As a result, lighter DM will lead to less recoil energy of particles in the Earth crust and suffer from less attenuation. 
It is noted that the total number of DM should be normalized in two scenarios related the Earth attenuation in order to verify the correctness in numerical calculations. 
Therefore, one cannot neglect the attenuation effect in sensitivity studies. As shown in the following section, we will see the ceiling limit caused by the Earth attenuation effect in re-analyzing experimental data.
\begin{figure}[!t]
\centering
\includegraphics[width=0.7\linewidth]{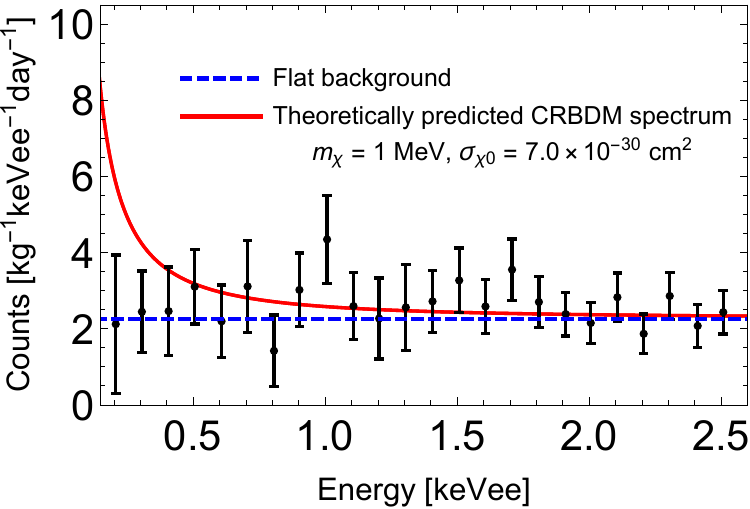}
\caption{The measured energy recoil spectrum in CDEX-10 with black data points plus error bars~\cite{CDEX10}, a flat background assumption in blue dashed lines the predicted CRBDM recoil spectrum in red solid line for $m_\chi = 1~\textrm{MeV}$ and $\sigma_{\chi0}=7\times10^{-30}$ cm$^2$. Here "eVee" represents electron equivalent energy derived from a charge calibration.}
\label{fig:expec}
\end{figure}

\subsection{Recoil spectrum}
\label{subsec:spectrum}
To obtain the recoil spectrum from CRBDM, we focus on the elastic and isotropic collision of DM and target particle in the detector.
By neglecting energy dependence, we get the recoil spectrum by integrating the kinetic energy of DM particle $dT_\chi$: 
\begin{equation}
\label{equ::rspectrum}
\frac{dR}{dE_R} = \frac{1}{m_N}
\int_{T_\chi^\mathrm{min}}^\infty  \frac{\sigma_{\chi N}}{E_R^{max}}
\frac{d\Phi_{\chi}}{dT_\chi^z} dT_\chi^z\,,
\end{equation}
where $T_\chi^z$ and $\frac{d\Phi_{\chi}}{dT_\chi^z} = \frac{dT_\chi}{dT_\chi^z} \frac{d\Phi_{\chi}}{dT_\chi}$ are the kinetic energy of DM and the DM flux in the detector with depth of $z$, respectively.
Here the $\sigma_{\chi N}$ is given by: $\sigma_{\chi N}=\sigma_{\chi N}^0 F^2(E_R)$ where the $F(E_R)$ is the form factor~\cite{SHM}. Since the momentum transfer is small in the process of DM-nucleus scattering, we adopt the Helm model which describes the nucleon distribution of the target nucleus with two parameters: the effective nuclear radius and the surface thickness~\cite{nuclearformfactor,PhysRev.104.1466}.
We will follow the similar expression in \forref{equ::min} and \forref{equ::relativty_relation} with $\theta=0$ to calculate $T_\chi^\mathrm{min}$ and $E_R^{max}$.
For scintillation and ionization detectors calibrated with $\gamma$ sources, the observable nuclear recoil energy $E_v$ differs from the true recoil energy $E_R$: $E_v = f_n(E_R) E_R = F(E_R)$, where $f_n$ is called the quenching factor as a function of $E_R$
calculated by the TRIM software package~\cite{ZIEGLER20041027, CDEX2014, Yang:2017} and $F(E_R)$ is more convenient to make the derivation.
Besides, because of a finite detector energy resolution and electronic noise, recoils at energy $E_R'$ would be observed as a Gaussian distribution.
Consequently, we will obtain the modified recoil spectrum: 
\begin{equation} 
\label{equ::correction}
\frac{dR}{dE_v'}=\frac{1}{\sqrt{2\pi}}\int{\frac{1}{\Delta E_v}\frac{dR}{dE_R} \frac{dE_R}{dE_v} \exp{[-\frac{(E_v'-E_v)^2}{2\Delta E_v^2}]}dE_v}
=
\frac{1}{\sqrt{2\pi}}\int{\frac{1}{\Delta E_v}\frac{dR}{dE_R} \frac{1}{dF/dE_R}[F^{-1}(E_v)] dE_v}
\,,
\end{equation}
where $\Delta E_v$ is the energy resolution.
As a demonstration, \figref{fig:expec} shows the energy recoil spectrum from CDEX-10 data, and the predicted recoil spectrum for DM with $m_\chi = 1$ MeV and $\sigma_{\chi 0}$. Note that the quenching factor and the energy resolution are included in the calculation of theoretical predictions. The effect of the detector efficiency has been included in the experimental data so that we can directly compare the experimental data with the theoretical expectation from~\forref{equ::correction}~\footnote{Private communications with the CDEX collaboration.}.

\subsection{Neutrino floor}
\label{subsec:vfloor}
Future DM direct detection experiments will encounter an irreducible background from solar, atmospheric and diffusive supernovae neutrinos through the CE$\nu$NS process~\cite{Strigari:2009bq, Billard:2013qya, Dutta:2019oaj, OHare:2021utq}, which will hinder our searches on CRBDM. It is a question in statistics whether DM direct detection experiments will see neutrino events and how to evaluate the impact of neutrino backgrounds on the DM detection. The neutrino floor is defined as the discovery limit: when the true DM model lies above the limit, a given experiment has a 90$\%$ probability to obtain more than $3\sigma$ DM signals.

CE$\nu$NS~\cite{PhysRevD.9.1389} is the dominant channel of neutrino backgrounds due to the coherent enhancement. The differential cross section of the neutrino-nucleus scattering are given by:
\begin{equation}
	\frac{d \sigma\left(E_{\nu}, E_{r}\right)}{d E_{r}}=\frac{G_{f}^{2}}{4 \pi} Q_{w}^{2} m_{N}\left(1-\frac{m_{N} E_{r}}{2 E_{\nu}^{2}}\right) F^{2}\left(E_{r}\right)
	\,,
\end{equation}
where $Q_{W}=N-\left(1-4 \sin \theta_{W}^{2}\right) Z$ is the weak charge, $\theta_{W}$ is the weak mixing angle, and $F\left(E_{r}\right)$ is the Helm form factor. By analogy with \forref{equ::rspectrum}, one can obtain the CE$\nu$NS event rates in a DM detector. Then we construct a likelihood function $\mathscr{L}\left(\sigma_{\chi 0}, \vec{\phi}\right)$ based on the Monte Carlo data, where $\vec{\phi}$ stands for nuisance parameters including uncertainties of different neutrino sources. In order to obtain the discovery limit, we take the background-only case as the null hypothesis $H_0$ and the alternative hypothesis $H_1$ includes both DM signals and backgrounds. Thus, the test statistic $q_0$ can be written as following:
\begin{equation}
	q_{0}=\left\{\begin{array}{cc}
		-2 \ln \frac{\mathscr{L}\left(\sigma_{\chi 0}=0, \hat{\hat{\vec{\phi}}}\right)}{\mathscr{L}\left(\hat{\sigma}_{\chi 0}, \hat{\vec{\phi}}\right)} & \hat{\sigma}_{\chi 0}>0 \\
		0 & \hat{\sigma}_{\chi 0}<0
	\end{array}\right. \,,
\end{equation}
where $\hat{\hat{\vec{\phi}}}$ is the value of $\phi$ which maximizes $\mathscr{L}$ for $H_0$, while $\hat{\sigma}_{\chi 0} \text{ and }\hat{\vec{\phi}}$ globally maximize $\mathscr{L}$ for $H_1$. Because $q_0$ asymptotically follows a $\chi^2$ distribution with one degree of freedom according to the Wilk's theorem, the $3\sigma$ statistical significance means the test statistics of an observation $q_\mathrm{obs}$ larger or equal to 9. Finally, we carry out 500 times Monte Carlo experiments with the same $m_\chi$ and $\sigma_{\chi 0}$ and apply the statistical principle to obtain the neutrino floor. 

\section{Constraints with CDEX-10 data}
\label{sec:constraints}
\begin{figure}[!t]
\centering
\includegraphics[width=0.45\linewidth]{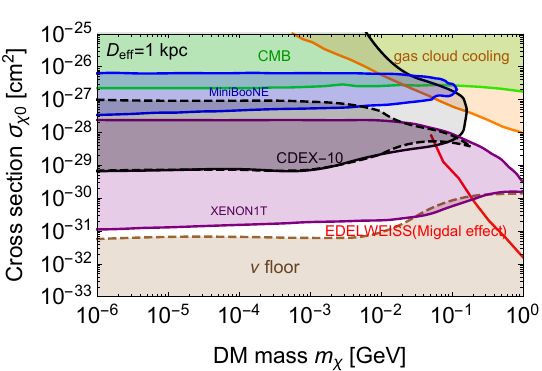}
\includegraphics[width=0.45\linewidth]{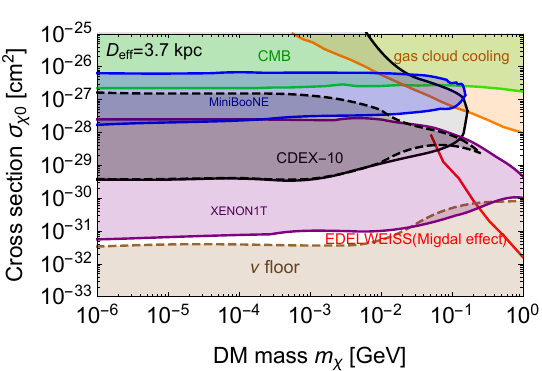}
\caption{\label{fig:crdbmexclusion}Constraints on spin-independent cross section of DM-nucleon interactions from CDEX-10 (gray), XENON1T (purple) and MiniBooNE (blue)~\cite{CRBDM} in the framework of CRBDM at a 90\% C.L. The gray region enclosed by solid and dashed black lines stand for constraint from CDEX-10 under the ballistic trajectory approximation with and without considering the form factor, respectively. As a comparison, we also include the exclusion limit from EDELWEISS~\cite{EDELWEISS} based on the Migdal effect and from CMB~\cite{Xu:2018efh} (green) as well as the gas cloud cooling~\cite{Bhoonah:2018wmw} (orange). The neutrino floor (brown) is obtained with an exposure of 1 ton$\cdot$year and a recoil energy threshold of 1 keV for the Germanium detector.}
\end{figure}
The CDEX-10 experiment located at China Jinping Underground Laboratory(CJPL) with about 2400 m of rock overburden makes use of p-type point contact Germanium (pPCGe) to reconstruct the ionization signal generated by the collision between DM and target particles~\cite{Yu-Cheng:2013iaa}. 
With the energy threshold as low as 160 eVee, CDEX-10 improves limits on $\sigma_{\chi 0}$ down to $m_\chi$ of 2 GeV.
%Therefore, with the data from germanium detector and under the assumption of CRBDM, there would be better constraints on the light DM.
To obtain the constraints, we adopt the minimum-$\chi^2$ analysis as the statistical method.
Considering that high energy $\gamma-$rays from ambient radioactivity produce flat electron-recoil backgrounds at low energy~\cite{CDEX2014,CDEX2016}, two free and positive parameters $\sigma_{\chi0}$ and b, which characterize the flat backgrounds and the spin-independent DM-nucleon scattering cross section $\sigma_{\chi0}$, are used to construct the statistical measure such as:
\begin{equation}
\label{equ::kfdef}
\chi^2(\sigma, b; \rho)\equiv\sum_k\frac{(S_k^{Th}(\sigma;\rho)+b-S_k^{Ex})^2}{\sigma_k^2}
	+ \frac{(\rho-\hat{\rho})^2}{(\sigma_\rho)^2} \,,~~~~
\chi^2(\sigma, b) = \min_\rho \chi^2(\sigma, b; \rho)	\,,
\end{equation}
where $S_k^{Ex}$ is the experimental data with the uncertainty $\sigma_k$, and $S_k^{Th}$ stands for the theoretical predictions in each bin, respectively.
Note that we add a pull term accounting for the astrophysical uncertainty, where $\rho$ is treated as a nuisance parameter with the central value $\hat{\rho} = 0.4~\textrm{GeV/cm}^3$ and the uncertainty $\sigma_\rho = 0.2~\textrm{GeV/cm}^3$.
To assess the exclusion limit, we calculate the probability in the likelihood analysis for background only data $H_b$ with respect to the null hypothesis $H_0=H_s+H_b$ where CRBDM is taken into account~\cite{Zyla:2020zbs}.
We assume that the data in each bin follow a Guassian distribution when we construct the likelihood function, and $\Delta \chi^2$ asymptotically follows a $\chi^2$ distribution with two degrees of freedom. The explicit distribution on $\Delta \chi^2$ could be obtained by Monte Carlo, which needs more inputs from the detector response beyond our current study.
In a likelihood analysis, the excluded parameter space is obtained by $\chi^2$ as follows:
\begin{equation}
\label{equ::kfdef2}
\chi^2(\sigma, b) \geq \chi^2_{min} + \Delta \chi^2 \,,
\end{equation}
where $\chi^2_{min}$ is the minimum value of $\chi^2$ and $\Delta \chi^2$ is associated with the number of considered parameters.
When the $\Delta \chi^2$ exceeds 4.61 for two degrees of freedom, we reject the hypothesis $H_0$ at a 90\% confidence level and obtain an exclusion limit in the parameter space. To be conservative, the DM mass and maximally allowed cross section to satisfy conditions in~\forref{equ::kfdef2} are represented as a point in the parameter space in \figref{sec:constraints}. Scanning over the interested low-mass DM range, we then figure out the exclusion limit for the spin-independent cross section as a lower part. Nevertheless, the Earth attenuation effect will dilute the strong exclusion limit since the bounced CRBDM might not be able to reach the detector threshold. This caveat will offer the upper ceiling limit. The eventual constraints on the DM parameter space will be obtained by a combination of exclusion limits for the lower part from an analysis of recoil spectrum in the DM direct detection experiment and the upper part from the Earth attentuation effect.

Following this strategy, we take the CDEX-10 experiment as a demonstration and reanalyze their data to obtain constraints on the DM mass $m_\chi$ and the cross section $\sigma_{\chi0}$ in the CRBDM hypothesis at a 90\% C.L. with two benchmark examples: $D_{\text{eff}}=1$ kpc and $D_{\text{eff}}=3.7$ kpc as shown in~\figref{fig:crdbmexclusion}. Compared with the Migdal effect, the CRBDM framework will reach a much lower mass region but suffers from the Earth attenuation effect. \zbl{However, cosmological constraints like the cosmic microwave background (CMB)~\cite{Xu:2018efh} and the gas cloud cooling~\cite{Bhoonah:2018wmw} help to exclude the parameter space above the reach of DM direct detection experiments.}
For the DM mass range 1 keV $<m_\chi <$ 1 MeV and $D_{\text{eff}}=1$ kpc, we reach the almost flat floor limit at $8.32\times10^{-30}$ cm$^2$ on spin-independent DM-nucleon scattering cross section, while this limit turns up for $m_\chi >1$ MeV and converges towards the ceiling limit caused by Earth attenuation. 
Here, we emphasize that the energy threshold also plays a crucial role for the attenuation effect. The low threshold of the CDEX detector somehow compensates for the sensitivity loss caused by the attenuation, though the CDEX experiment might suffer a stronger attenuation effect due to the thicker overburden. Therefore, the ceiling of the CDEX constraint can be higher than that of the XENON1T constraint.
As shown in~\figref{fig:crdbmexclusion}, the gray region enclosed by solid and dashed black lines stand for the constraints under the ballistic trajectory approximation with and without considering the form factor, respectively. We notice that the constraint without the form factor suppression in the attenuation calculations is more conservative, except for minor difference around the right endpoint. Next we come to the question Why the form factor helps to extend the ceiling limit. DM particles with higher energies suffer less energy loss when going through the crust due to the form factor suppression. Then more DM particles can reach the underground detector and the attenuation effect is relieved.
The current result from CDEX-10 is one order of magnitude better than the neutrino experiment MiniBooNE but not comparable to the XENON1T data. Due to the attenuation, there will be a narrow gap for the excluded parameter regions from MiniBooNE and XENON1T. 
Given a larger $D_{\text{eff}}$, the floor limit for each experiment will become stronger, while the ceiling limit induced from the Earth attenuation will hardly make changes. In a word, our analysis is more conservative compared with the previous study. The effective distance can enlarge the excluded regions for MiniBooNE and XENON1T to narrow the gap or even close it. It is then valuable to make a further study on $D_{\text{eff}}$ in detail, especially for the uncertain CR spatial distribution. With the current CRBDM hypothesis, the CDEX-10 result can close the gap in a nice manner. This fact highlights the importance of multiple detection technologies and combined analysis to examine the CRBDM mechanism.

Distinct from the ballistic trajectory assumption we take in deriving the attenuation effect, several studies performed a Monte Carlo (MC) simulation on the scattering process in the crust in detail~\cite{Xia:2021vbz, Xu:2022aeo, CDEX:2021cll,PROSPECT:2021awi,PandaX:2021kai}. The MC simulation is able to calculate the deflection of DM going through the Earth crust, which results in fewer DM particles reaching the underground detector. Thus, compared with the ballistic trajectory approach considering the form factor in the attenuation calculation, the MC simulation provides a slightly more conservative constraint. In addition, in the ballistic approach without considering the form factor, there is a suppression at the higher energy range ($\sim 0.1$ GeV). Due to the suppression, DM can not generate enough recoil energy and be detected when the cross section climbs over the ceiling limit. \zblb{Since DM with higher energies can go through the crust for the form factor suppression, both the MC method and the ballistic approach in this case lead to too many recoil events. Thus, in order to be consistent with the experimental result, it is needed to enhance the attenuation effect in terms of a larger DM cross section.}

As for the smaller DM cross section, the floor limit suffers less from the Earth attenuation compared with the ceiling bound. Therefore, the differences among the floor limits derived from various approaches should be smaller than that of the ceiling limits. However, for the larger cross section or the thicker crust leading to a stronger attenuation effect, the differences become larger and more careful treatment is needed to calculate the Earth attenuation. See Appendix A for more details. The data analyses in different experiments based on the MC simulation sets complementary constraints on the parameter space and the global floor limit comes down to about $3 \times 10^{-32}$ cm$^2$ on spin-independent DM-nucleon scattering cross section for the DM mass below 0.1~GeV. For the upper bound, the constraint derived from the MC simulation can exclude the parameter space between the floor limit and excluded regions from CMB and the gas cloud cooling constraints.

Coherent neutrino-neucleus scatterings will become intrinsic backgrounds of future DM direct detection experiments as soon as they make non-negligible contributions to nuclear recoils in the detector. Compared to the Xenon-based experiment, Germanium detector can obtain more energy transferred from neutrinos with tiny masses and has a lower energy threshold, which leads to more neutrino backgrounds. Taking neutrino backgrounds into account, we calculate the sensitivity of future germanium detectors on CRBDM with an exposure of about 1 ton$\cdot$year and an energy threshold as low as 160 eVee in case of CDEX upgrades. The corresponding result of the neutrino floor is shown at the bottom area in ~\figref{fig:crdbmexclusion}. For the DM mass range 1 keV $<m_\chi <$ 1 MeV and $D_{\text{eff}}=1$ kpc, it is found that we reach the neutrino floor at the cross section of about $3 \times10^{-32}$ cm$^2$ in future Germanium-based DM experiments. 
In other words, the large-scale Germanium-based detector with a lower threshold might observe CE$\nu$NS events and simultaneously facilitate the study of astrophysical neutrino physics.

\section{Conclusions}
\label{sec:conclusion}
Given that DM could elastically scatter with target nuclei in detectors, it is straightforward to think of the collisions between relativistic cosmic rays and low-speed DM in the galaxy. In this way, CRBDM would become more energetic to make the recoil energy above the threshold in detectors, especially for DM in the low mass range. Previous studies were conducted to constrain CRBDM for various experiments including MiniBooNE and XENON1T. There exists a gap in the excluded parameter regions between MiniBooNE and XENON1T partially due to the fact that the Earth attenuation effect will dilute the experimental constraints. Such a gap strongly depends on the effective distance determined by CR models where we need further investigations on the CR spatial distribution. We have reviewed the theoretical framework for the CRBDM, including the modified DM fluxes, the Earth attenuation effect from the overburden and the energy recoil spectrum for the general detector target nuclei. Taking the CDEX-10 experiment into account, we have re-analyzed 102.8 kg$\times$day data to constrain CRBDM, and obtained constraints on the spin-independent DM-nucleon cross section in the low mass range 1~keV~$<m_\chi <$~1 GeV, surpassing the detector energy threshold limitation in a classical analysis. Migdal effect can also lead to constraining the light DM but is still not comparable to the CRBDM mechanism. We have found that the CDEX-10 result is able to close the gap unambiguiously regardless of the assumed effective distance in the parameter space between MiniBooNE and XENON1T constraints.

For future direct detection experiments, the same strategy to constrain CRBDM could be applied. Note that it is still necessary to take careful treatment of uncertainties from astrophysical and nuclear physics inputs as well as the detector response in order to give more rigorous results on CRBDM. In addition, we also quantitiviely provide the neutrino floor on searching for CRBDM in future direct detection experiments using Germanium. Our work should encourage the experimental collaborations to press forward with development of the next generation Germanium detector and joint analysis with other experiments using different detection technologies. Recently, the XENON1T collaboration reported the low-energy ER excess~\cite{Aprile:2020tmw}. CRBDM might be able to explain the observed ER excess in the LXe detector~\cite{Cao:2020bwd}. We look forward to further examination by next-generation low-threshold DM direct detection experiments. 

\section*{Acknowledgement}
We appreciate Dr. Li-Tao Yang for communications and useful discussions. Thanks Dr. Cheng-Cheng Han and Dr. Sampsa Vihonen for careful reading of the manuscript. This work was supported in part by Guangdong Basic and Applied Basic Research Foundation under Grant No. 2019A1515012216, National Undergraduate Innovation and Entrepreneurship Training Program No. 20201023, and the CAS Center for Excellence in Particle Physics (CCEPP).

\section*{Appendix A: Attenuation Effect for Floor Limits}
\begin{figure}[!t]
	\centering
	\includegraphics[width=0.9\linewidth]{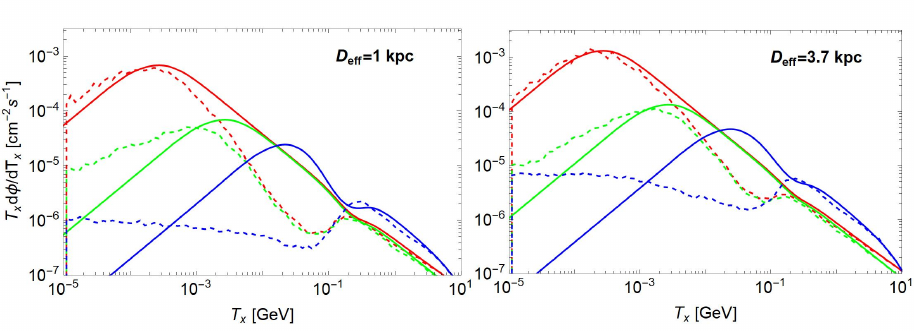}
	\includegraphics[width=0.9\linewidth]{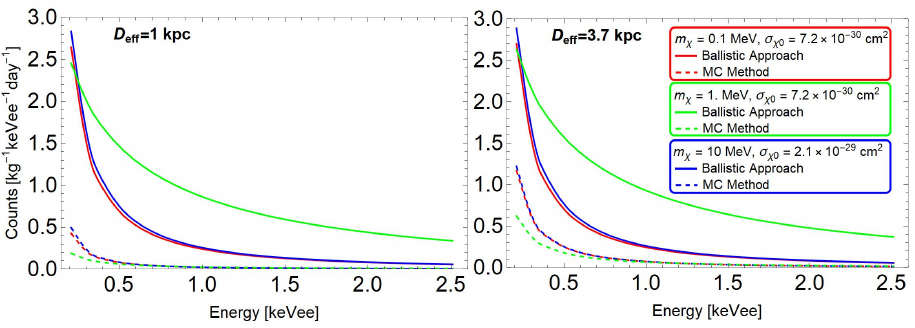}
	\caption{\label{fig:appendix}The CRBDM fluxes and theoretical spectra derived from the ballistic approach (solid line) and the MC method (dashed line). \zblb{Form factor effects are included here in all panels.}}
\end{figure}
To explicitly present the differences in calculating the attenuation effect with the ballistic approach and the MC method, we choose three benchmark points located on the floor limits in \figref{fig:crdbmexclusion} and show the corresponding DM fluxes and recoil spectra as shown in \figref{fig:appendix}. A simulation code Darkprop is utilized for calculations of the attenuation  effect~\cite{Xia:2021vbz}. 
From the upper part of \figref{fig:appendix}, one can see that the fluxes in the MC case have a tail distribution in the higher energy range due to the form factor suppression. As for the smaller cross section or DM mass, the attenuation effect gets weaker so that the differences between the ballistic approach and the MC method get diminished eventually.
From the bottom of \figref{fig:appendix}, we find that the recoil spectrum of the MC method is a bit more conservative than that of the ballistic approach for the longer effective distance with $D_\text{eff}=3.7~\text{kpc}$. Note that whether the form factor is considered in the ballistic approach or not, the recoil spectra are almost the same. For the closer collision location with $D_\text{eff}=1~\text{kpc}$ and a larger cross section, the spectrum discrepancy gets increased in a comparison of the MC method and ballistic approach. Consequently, the floor limit would be lifted if the MC method is adopted, while a larger $D_\text{eff}$ can dilute the relevant modification. 

The larger cross section or the thicker crust would lead to a stronger attenuation effect. Thus, we strongly advise that a more careful treatment is needed to judge the Earth attenuation effect.

%For $D_\text{eff}=1~\text{kpc}$, the red, green and blue lines stands for $\{m_\chi, \sigma_{\chi0}\}=\{0.1~\text{MeV}, 7.2\times10^{-30} \text{cm}^2\},\{1~\text{MeV}, 7.2\times10^{-30} \text{cm}^2\} \text{and} \{10~\text{MeV}, 2.1\times10^{-29} \text{cm}^2\}$, respectively. For $D_\text{eff}=3.7~\text{kpc}$, the red, green and blue lines stands for $\{m_\chi, \sigma_{\chi0}\}=\{0.1~\text{MeV}, 3.8\times10^{-30} \text{cm}^2\},\{1~\text{MeV}, 3.8\times10^{-30} \text{cm}^2\} \text{and} \{10~\text{MeV}, 1.2\times10^{-29} \text{cm}^2\}$, respectively.

%\bibliographystyle{unsrt}	
\bibliography{reference}	
	
\end{document}